\documentclass[preprint,aps,nofootinbib]{revtex4}

\usepackage{dcolumn}
\usepackage{bm}
\usepackage{epsfig}
\usepackage{graphicx}
\usepackage[bookmarks=true]{hyperref}
\usepackage[caption=false]{subfig}

\usepackage{amsmath}
\usepackage{amsfonts}
\usepackage{amssymb}

\usepackage{amsmath}
\usepackage{amsfonts}
\usepackage{amssymb}
\usepackage{color}

\def\be{\begin{equation}}
\def\ee{\end{equation}}
\def\bea{\begin{eqnarray}}
\def\eea{\end{eqnarray}}

\newcommand{\h}[1]{\hat{#1}}

\def\slashchar#1{\setbox0=\hbox{$#1$}           
   \dimen0=\wd0                                 
   \setbox1=\hbox{/} \dimen1=\wd1               
   \ifdim\dimen0>\dimen1                        
      \rlap{\hbox to \dimen0{\hfil/\hfil}}      
      #1                                        
   \else                                        
      \rlap{\hbox to \dimen1{\hfil$#1$\hfil}}   
      /                                         
   \fi}

\begin{document}

\title{A Redux on ``When is the Top Quark a Parton?''}
\vspace*{1cm}

\author{\vspace{1cm} S.~Dawson$^{\, a}$, A.~Ismail$^{\, b,c}$ and Ian Low$^{\, b,d}$ }

\affiliation{
\vspace*{.5cm}
  \mbox{$^a$ Physics Department, Brookhaven National Laboratory, Upton, NY 11973}\\
 \mbox{$^b$ High Energy Physics Division, Argonne National Laboratory, Argonne, IL 60439}\\
\mbox{$^c$ Department of Physics, University of Illinois, Chicago, IL 60607}\\
\mbox{$^d$ Department of Physics and Astronomy, Northwestern University, Evanston, IL 60208} \\
\vspace*{1cm}}

\begin{abstract}
\vspace*{0.5cm}
If a new heavy particle $\phi$ is produced in  association with the top quark in a hadron collider, the production cross section exhibits a collinear singularity of the form  $\log (m_\phi/m_t)$, which can be resummed by introducing a top quark parton distribution function (PDF). We reassess the necessity of such resummation in the context of a
high energy $pp$ collider. We find that the introduction of a top PDF typically has a small effect at $\sqrt{S}\sim  100~\mathrm{TeV}$  due to three factors: 1) $\alpha_s$ at the scale $\mu=m_\phi$ is quite small  when $\log (m_\phi/m_t)$ is large, 2) the Bjorken $x\ll 1$ for $m_\phi \alt 10$ TeV, and 3) the kinematic region where $\log (m_\phi/m_t) \gg 1$ is suppressed by phase space. We consider the example of $pp\rightarrow t H^+$ at next-to-leading logarithm (NLL) order and show that, in terms of the total cross section, the effect of a top PDF is generically smaller than that of a bottom PDF in the associated production of $b\phi$.  However, in the $p_T$ distribution of the charged Higgs, the NLL calculation using a top PDF is crucial to generate the $p_T$ distribution for $p_T\alt m_t$.
\end{abstract}

\maketitle

\section{Introduction}

The production of heavy quarks in a hadronic scattering process is interesting because it involves several hard scales.  The
question of whether the bottom quark is appropriately treated as a parton at the LHC has received significant theoretical
attention~\cite{Dicus:1998hs,Campbell:2002zm,Maltoni:2003pn,Dittmaier:2003ej,Dawson:2003kb,Campbell:2004pu,Dawson:2004sh,Maltoni:2005wd,Dawson:2005vi,Maltoni:2012pa}.  
A particularly noteworthy example is that of Higgs production in association with $b$ quarks, where at scales $Q \ll m_b$, one can perform the calculation in a 4-flavor number scheme (FNS), with the lowest order
process being $gg\rightarrow b \bar{b} H$.  In this scheme, the $b$ quark mass is included exactly in the final
state kinematics.
However, for scales $Q \gg m_b$, there are large logarithms of 
the form $\log (Q^2/m_b^2)$~\cite{Aivazis:1993kh,Aivazis:1993pi,Dicus:1998hs,Kniehl:2005mk}. These logarithms can potentially spoil the convergence of a fixed order perturbative
calculation.
Typically, the issue of large logarithms is rectified by resumming these logarithms into a $b$ quark PDF, leading to a 5FNS, in
which incoming $b$ quarks are treated as
massless partons~\cite{Barnett:1987jw,Olness:1987ep,Aivazis:1993kh,Aivazis:1993pi,Olness:1997yc,Kramer:2000hn}. The 4- and 5-FNS PDF schemes represent alternative ways of organizing perturbation theory, and a 
correct treatment should interpolate between the two schemes in the appropriate kinematic
regimes~\cite{Kusina:2013qra}. If we could calculate to all orders
in $\alpha_s$, the results of the different schemes would be identical. For processes involving the production of $b$ quarks, the calculations
in the $5$FNS are simpler, while the calculations in the $4$FNS include the kinematics of the outgoing $b$ 
quark at lowest order.  For the Large Hadron Collider (LHC), it has been demonstrated that
consistent results for both the total cross section and kinematic distributions 
for Higgs production in association with $b$ quarks can be obtained in both 
PDF schemes~\cite{Dawson:2003kb,Campbell:2004pu,Dittmaier:2011ti,Harlander:2011aa,Dittmaier:2012vm}.  

In this paper, we examine the question of whether the
top quark should be treated as a parton at high center-of-mass energy,
which corresponds to a 6FNS.  This question was originally considered in
the pioneering works of Refs.~\cite{Barnett:1987jw,Olness:1987ep}, which predate the discovery of the top quark.
We re-examine the question in light of our knowledge on the top mass as well as a potential $\sqrt{S}\sim 100~\mathrm{TeV}$ $pp$ collider. We evaluate the impact of resumming collinear logarithms involving the $t$ quark at scales that would be accessible at such a collider, testing the efficacy of using a top PDF. Additionally, we compare with the case of lighter quarks at lower collider energies. Our results are generically applicable to the production of heavy particles in association with $t$ quarks at hadron colliders.

Now, in the Standard Model, the rate for Higgs boson production in association with a top quark is quite 
small~\cite{Beenakker:2001rj,Reina:2001sf,Reina:2001bc,Beenakker:2002nc,Dawson:2002tg,Dawson:2003zu}, and there are no large logarithms to be resummed.
In theories with extra Higgs multiplets, however, the cross section for heavy Higgs production in association with top quarks may be significant.
For instance, in type II two Higgs doublet models such as the Minimal Supersymmetric Standard Model (MSSM), heavy Higgs production can be enhanced for small values of $\tan
\beta$~\cite{Plehn:2002vy,Boos:2003yi,Berger:2003sm,Dittmaier:2009np,Weydert:2009vr}. As a concrete example of the relevance of the top PDF, we consider
charged Higgs production in association with a top quark, although it is worth mentioning that the case of heavy neutral Higgs production in association with top quarks can be studied in a similar fashion. 
Charged Higgs production  has been studied in the past \cite{Olness:1987ep,Barnett:1987jw}, and our contribution is to 
discuss the new features which arise at partonic energies much larger than the top quark
mass, $\sqrt{\h{s}} \gg m_t$.
In the 5FNS, where the top quark is not treated as a parton, the leading order process is $g \bar{b} \rightarrow \bar{t} H^+$, while in the 6FNS it is 
$t {\bar{b}} \rightarrow H^+$.
We demonstrate the effects of the collinear logarithms of the form $\log (Q^2/m_t^2)$ in the 6FNS and compare
 to the 5FNS. The NNPDF collaboration~\cite{Ball:2011mu,Ball:2011uy,Ball:2012cx} has produced a set of 6FNS PDFs,
which allows  for a quantitative analysis.
We present both total and differential cross sections, showing the effect of the top quark PDF resummation of collinear logarithms for large charged
Higgs masses.

In Section~\ref{sec:counting}, we review the organization of perturbation theory in schemes with different numbers of flavors, describing how collinear logarithms may be resummed into heavy quark PDFs. Next, Section~\ref{sec:3facs} contains an exploration of the quantitative effect of this resummation. We examine the variation of its numerical impact with heavy quark mass and collider energy, considering the impact of the phase space of the collinear logarithm as well. Section~\ref{sec:charged} details the calculation of the cross section for charged Higgs production in association with a single top quark at leading logarithm (LL) and next-to-leading logarithm (NLL). We compare our 6FNS results to the 5FNS calculation at leading order in $\alpha_s$. Our conclusions are in Section~\ref{sec:conclusion}.

\section{Counting Logarithms and {\large $\alpha_{\lowercase{s}}$}}
\label{sec:counting}
Production of a new heavy particle $\phi$ in association with heavy quarks,\footnote{We define heavy quarks to be those whose masses are large enough for the running strong coupling $\alpha_s(m_{q})$ to stay in the perturbative regime. Therefore, the top and bottom quarks are considered heavy quarks, while the charm quark is a borderline case. Furthermore, we are interested in scenarios where $m_\phi \gg m_q$.} $q_h$, is a nice illustration of multi-scale processes in quantum field theory.  In the presence of two distinct scales, $m_\phi$ and $m_q$, perturbative calculations exhibit potentially large logarithms $\log(m_\phi/m_q)$ and power corrections in 
$m_q^2/m_\phi^2$. When $m_\phi \gg m_q$, power corrections become less important while the large logarithms could potentially spoil the perturbative expansion in the coupling constant~\cite{Dicus:1998hs}. In particular, because the heavy quarks are much heavier than the proton, it is easy to trace the origin of the logarithms to the process of a gluon $g$ splitting into a $q_h\bar{q}_h$ pair inside the proton~\cite{Barnett:1987jw,Olness:1987ep}:
\be
g(p) \to q_h(k_q) + \bar{q}_h({k_{\bar{q}}}) \ .
\ee
Obviously an on-shell massless particle cannot decay into two massive particles that are both on-shell, because otherwise the rest frame of the two massive particles would define a rest frame for the gluon, which does not exist. One could, however, consider the kinematic region where only two particles, for example $g$ and $q_h$, are on-shell, in which case $\bar{q}_h$ cannot be an external state and must be an internal line with the 
propagator~\cite{Altarelli:1979ub,Harris:2001sx}
\be
\frac{1}{(p-k_q)^2-m_q^2} = -\frac{1}{  2 p\cdot k_q} \ .
\ee
If we go to a frame where
\be
p=(E, 0, 0, E) \ , \quad
 k_q=(\sqrt{m_q^2 + |\vec{k}_q|^2} , |\vec{k}_q| \sin\theta, 0,  |\vec{k}_q| \cos\theta) \ ,  \quad k_{\bar{q}} = p-k_q \ ,
\ee
the denominator of the propagator for $\bar{q}$ is
\be
2p\cdot k_q = 2E |\vec{k}_q|  \left(\sqrt{1+ \frac{m_q^2}{ |\vec{k}_q|^2}} - \cos\theta\right) \ ,
\ee
which never vanishes unless $m_q=0$ and $\cos\theta=1$. This is the famous collinear singularity in the three-body kinematics, which we see explicitly is regulated by the non-zero quark mass~\cite{Collins:1985ue}. 
Upon integrating over the phase space, the collinear singularity gives rise to the factor $\log (Q^2/m_q^2)$, where $Q^2$ is the typical hard momentum transfer in the process.\footnote{There is a subtlety involving whether the splitting gluon is in the initial state or the final state. In this work we are interested in the initial state logarithms, as the final state 
logarithms can be cancelled by defining sufficiently inclusive observables or resummed by introducing a fragmentation function.} For the production of a new heavy particle $\phi$, we expect $Q^2 \sim m_\phi^2$. However, it is important to emphasize that this is only an order-of-magnitude estimate.

The existence of potentially large logarithms suggests the necessity to re-organize the perturbative expansion. To achieve this goal, it is conceptually clearest to introduce an effective theory where the heavy quarks are treated as light degrees of freedom when the typical hard scale in the process satisfies $Q^2 \gg m_q^2$. On the other hand, when $Q^2\ll m_q^2$, the heavy quarks are treated as genuine heavy degrees of freedom. This subject has a long history~\cite{Witten:1975bh,Collins:1978wz}, and in the present context it was first discussed in Refs.~\cite{Olness:1987ep,Barnett:1987jw}. In particular, the approach where the heavy quark is considered ``heavy,'' in the sense that it is not a constituent of the proton, is called the $(n_f-1)$ FNS, where $n_f=4, 5$, and $6$ for the charm, bottom, and top quarks, respectively. On the other hand, in the  $n_f$ FNS the heavy quark is treated as a ``light'' parton inside the proton.

\begin{figure}[t]
\includegraphics[scale=0.6, angle=0]{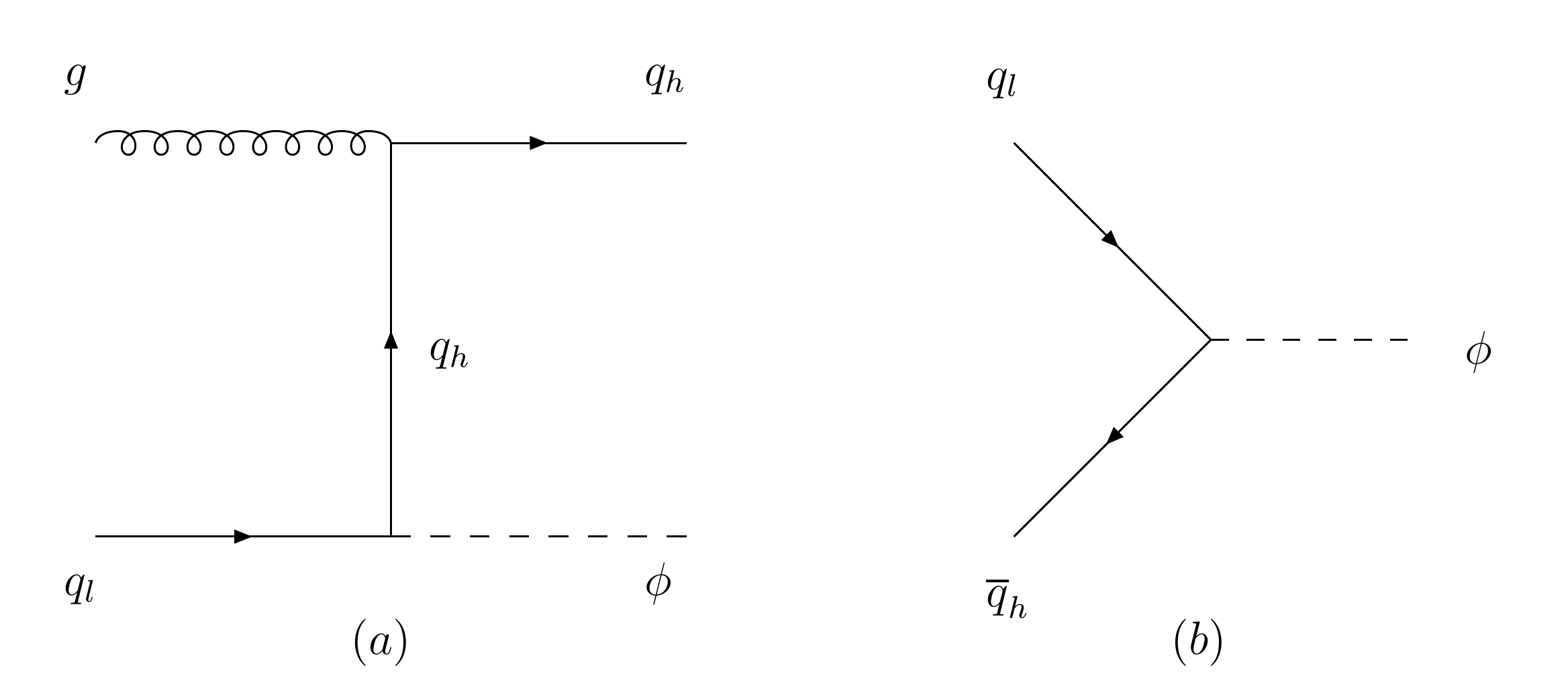}
\caption{\label{fig1}{\label{fg:fig1}(a) Feynman diagram for $gq_l\to q_h \phi $, (b) Feynman diagram for $q_l {\bar{q}}_h\to \phi$. 
}}
\end{figure}

In the $(n_f-1)$ FNS, the heavy quark never appears as an initial state, and the leading order (LO) process for the associated production is given by
\be
\label{eq:nf1lo}
g + q_l \to \phi + q_h\ ,
\ee
where $q_l$ represents a light  constituent of the proton. A representative Feynman diagram is shown in Fig.~\ref{fg:fig1}a. The cross section in perturbative QCD has the following series expansion at each order
in $\alpha_s$:
\bea
\sigma_{n_f-1} &\sim&\phantom{+}c_{11}\ \alpha_s L\phantom{^2}+ c_{12}\ \alpha_s  \nonumber \\
                           & & +\ c_{21}\ \alpha_s^2 L^2 + c_{22}\ \alpha_s^2 L\phantom{^2}  + c_{23}\ \alpha_s^2  \nonumber \\
                           && +\ c_{31}\ \alpha_s^3 L^3 + c_{32}\ \alpha_s^3 L^2 +c_{33}\  \alpha_s^3 L+ c_{34}\ \alpha_s^3 \nonumber \\
                           && +\ \cdots \ ,
\label{eq:signf1}
\eea                           
where $L\equiv\log (Q^2/m_q^2)$. It is then apparent that when $\alpha_s L \sim {\cal O}(1)$, the perturbative expansion in the $(n_f-1)$ FNS may be spoiled.

The dynamical origin of the logarithm
 comes from the collinear region where both heavy quarks from the gluon splitting are (approximately) collinear with the incoming gluon, in which case the heavy quark produced in association with the new particle $\phi$ simply goes down the beampipe, along with the remnants of the proton, and cannot be detected. Therefore, in this region of phase space, one should really think of the heavy quark as part of the proton, {\em i.e.} a parton inside the proton. This picture motivates the $n_f$ FNS where the heavy quark is considered as a parton inside the proton and a PDF is introduced. The logarithms
  in Eq.~(\ref{eq:signf1}) are then resummed via the Dokshitzer-Gribov-Lipatov-Altrarelli-Parisi (DGLAP)
   equations~\cite{Altarelli:1977zs} to all orders in $\alpha_s$, effectively re-organizing the perturbative expansion. When computing the heavy quark PDF, $f_q(x,\mu)$, using the one-loop DGLAP evolution, all $c_{n1}$ terms, $n\ge 1$,  in Eq.~(\ref{eq:signf1}) of the form $(\alpha_s L)^n$ are resummed into $f_q(x,\mu)$. This is the LL approximation. At two-loop evolution, in addition to $c_{n1}$, part of  the $c_{n2}$, $n\ge 2$, terms in Eq.~(\ref{eq:signf1}), are also resummed into the top PDF.
  The $(n_f-1)$ FNS and $n_f$ FNS are matched at the scale $\mu=m_q$, where
\be
f_q(x,m_q)=0 \ .
\ee
In this picture, the heavy quark can  be an initial state particle and the LO process for the production of $\phi$ is 
\be
\label{eq:nflo}
q_l + \bar{q}_h \to \phi \ ,
\ee
which is shown in Fig.~\ref{fg:fig1}b. Again, it is worth emphasizing the process 
$q_l {\bar{q}_h}\to \phi$ in the $n_f$ FNS is nothing but the 
$g  {q_l} \to q_h\phi$ process in the $(n_f-1)$ FNS when the final state $q_h$ is collinear with $g$ and has a small $p_T$, thereby escaping detection. To account for all terms proportional to $\alpha_s (\alpha_s L)^n$ at NLL accuracy, one would need to include ${\cal O}(\alpha_s)$ corrections to Eq.~(\ref{eq:nflo}) as well as new processes to be specified later.

It is instructive to consider approximate solutions of the DGLAP evolution, truncated at finite orders in $\alpha_s$, where only a finite number of the logarithms 
are included. For example, at LO  and NLO in $\alpha_s$, the 1-loop and 2-loop approximate heavy quark PDFs in the $n_f$ FNS are given by
\bea
\label{eq:1looppdf}
\tilde{f}^{(1)} _q(x,\mu)&=& \frac{\alpha_s}{2\pi} \log\frac{\mu^2}{m_q^2} \int_x^1 \frac{dz}z\ P_{qg}(z)\ f_g(x/z,\mu) \ , \\
\tilde{f}^{(2)} _q(x,\mu)&=& \tilde{f}^{(1)} _q(x,\mu)  + \left(\frac{\alpha_s}{4\pi}\right)^2 \int_x^1 \frac{dz}z\ \Sigma_{n_f-1}(x/z,\mu)\ a_{\Sigma,q}(z,\mu^2/m_q^2) \nonumber \\
&& + \left(\frac{\alpha_s}{4\pi}\right)^2  \int_x^1 \frac{dz}z\ f_g(x/z,\mu) \ a_{g,q}(z,\mu^2/m_q^2) \ ,
\label{eq:2looppdf}
\eea
where  $f_g(x,\mu)$ and $\Sigma_{n_f-1}(x,\mu) = \sum_{i=1}^{n_f-1} ( f_{q_i} + f_{\bar{q}_i})$ are the gluon and the singlet PDFs, respectively, computed to the corresponding order in  $\alpha_s=\alpha_s(\mu)$. The LO gluon splitting function is well-known \cite{Altarelli:1977zs}:
\be
\label{eq:pqg}
P_{qg}(z) = \frac12 \left[ z^2 + (1-z)^2 \right] \ ,
\ee
while the two-loop coefficient functions are computed in Refs.~\cite{Buza:1996wv,Buza:1995ie} and collected in the appendix of Ref.~\cite{Maltoni:2012pa}, whose notation we follow. Schematically, the two-loop coefficients have the form
\be
a_{\Sigma,q}(z,\mu^2/m_q^2),\quad a_{g,q}(z,\mu^2/m_q^2) \quad \sim \quad \log^2 \frac{\mu^2}{m_q^2} \  +\  \log \frac{\mu^2}{m_q^2}  \ ,
\ee
where the coefficients of the logarithms are $z$-dependent. We see that $\tilde{f}^{(1)} _q$ captures the $c_{11}$ contribution in Eq.~(\ref{eq:signf1}), which is included in a LO computation in the $(n_f-1)$ FNS, 
 while $\tilde{f}^{(2)} _q$ contains $c_{11}$, $c_{21}$ and parts of the $c_{22}$ pieces.

If one were able to compute the cross section to all orders in perturbation theory, then the $(n_f-1)$ FNS and $n_f$ FNS would  give the same answer. However, the expansion parameters in the two schemes are different and, when truncated at finite order,  result in 
numerically
different cross sections. More specifically, the $(n_f-1)$ FNS is a series expansion in $\alpha_s$, while the $n_f$ FNS
is also an expansion in $L^{-1}$, since terms of the forms $(\alpha_s L)^n$, $\alpha_s (\alpha_s L)^{n-1}$, etc., are resummed at successive orders. This power counting is the same as that in single top 
production \cite{Stelzer:1997ns} and in Higgs production in association with bottom quarks~\cite{Dicus:1998hs}
in the $5$FNS using $b$ PDFs.
The LO processes in the $(n_f-1)$ and $n_f$ FNS for
$\phi$ production in association with a top quark
are given by $g q_l\to q_h\phi$ and 
$q_l\bar{q}_h\to \phi$, respectively, and contain the following contributions:
\bea
{\rm LO}_{n_f-1}&:&\phantom{+}   c_{11}\ \alpha_s L\phantom{^2}+ c_{12}\ \alpha_s \ ,\\
{\rm LO}_{n_f}&:&\phantom{+}   \sum_{n=1}^\infty c_{n1} ( \alpha_s L)^n \ .
\eea
 The calculation at LO in the $n_f$ FNS, which only involves a 2-to-1 process, is simpler than that in the $n_f-1$ FNS, which is a 2-to-2 process, and represents the LL approximation to the full cross section. However, the 2-to-1 process is clearly inadequate if the heavy quark in the final state has a significant transverse momentum $p_T$.

We work to NLL in the $n_f$ FNS, to include the effects of finite $p_T$ not present in the LL approximation.  To NLL, 
one computes the virtual and real corrections to $q_l \bar{q}_h \to \phi$. In the $n_f$ FNS, an NLL 
calculation requires not only the virtual and real corrections to $q_l \bar{q}_h\to \phi$, 
as well as the NLO evolution of the heavy quark PDF using DGLAP equations, but also
 the addition of the
processes $gq_l\to q_h\phi$ and $g \bar{q}_h \to \bar{q}_l \phi$, 
which now open up as new channels at this order. The latter process contributes only terms that are $\mathcal{O}(\alpha_s^2 L)$ and higher, with no terms proportional to $\mathcal{O}(\alpha_s)$ term as in the former process. Additionally, there is a subtlety in incorporating these new processes. Note that the 
$gq_l\to q_h\phi$ process contains, in addition to the $c_{21} \alpha_s$ contribution to the cross section, the $c_{11} \alpha_s L$ term that has already been resummed into the heavy quark PDF at LO$_{n_f}$. 
Therefore, na{\"\i}vely adding 
the contribution of $gq_l\to q_h\phi$ to the  LO$_{n_f}$ result 
would result in a double counting of 
the
$c_{11}$ term in the $n_f$ FNS.
This double counting
needs to be subtracted properly \cite{Olness:1987ep,Barnett:1987jw} by using the 1-loop approximated PDF in Eq.~(\ref{eq:1looppdf}). Once this is done, the remaining component of the $gq_l\to \phi q_h$ subprocess is only $\mathcal{O}(\alpha_s)$ and down by $L^{-1}$ when compared with $q_l \bar{q}_h\to \phi$.
In the end, the NLL result in the $n_f$ FNS contains the desired terms,
\bea
{\rm NLL}_{n_f}&:&\phantom{+}   \sum_{n=1}^\infty c_{n1} ( \alpha_s L)^n  \nonumber \\
                             && + \sum_{n=1}^\infty c_{n2} \ \alpha_s ( \alpha_s L)^{n-1}  \ .
\eea
In the above, the $c_{12}$ term comes from the subtracted $gq_l\to q_h \phi $ subprocess in the $n_f$ FNS; $c_{22}$ is obtained from the NLO PDF, $g \bar{q}_h \to \bar{q}_l \phi$ and the $\alpha_s$ correction to the $q_l{\overline q}_h\to \phi$ process; and the $c_{n2}$,  $n\ge 3$, terms are reproduced in the NLO heavy quark PDF.

\section{The Three Factors}
\label{sec:3facs}
In this section we discuss the three factors determining the importance of the collinear logarithms that are resummed into the heavy quark PDFs. 
\begin{figure}[t]
\includegraphics[scale=0.6, angle=0]{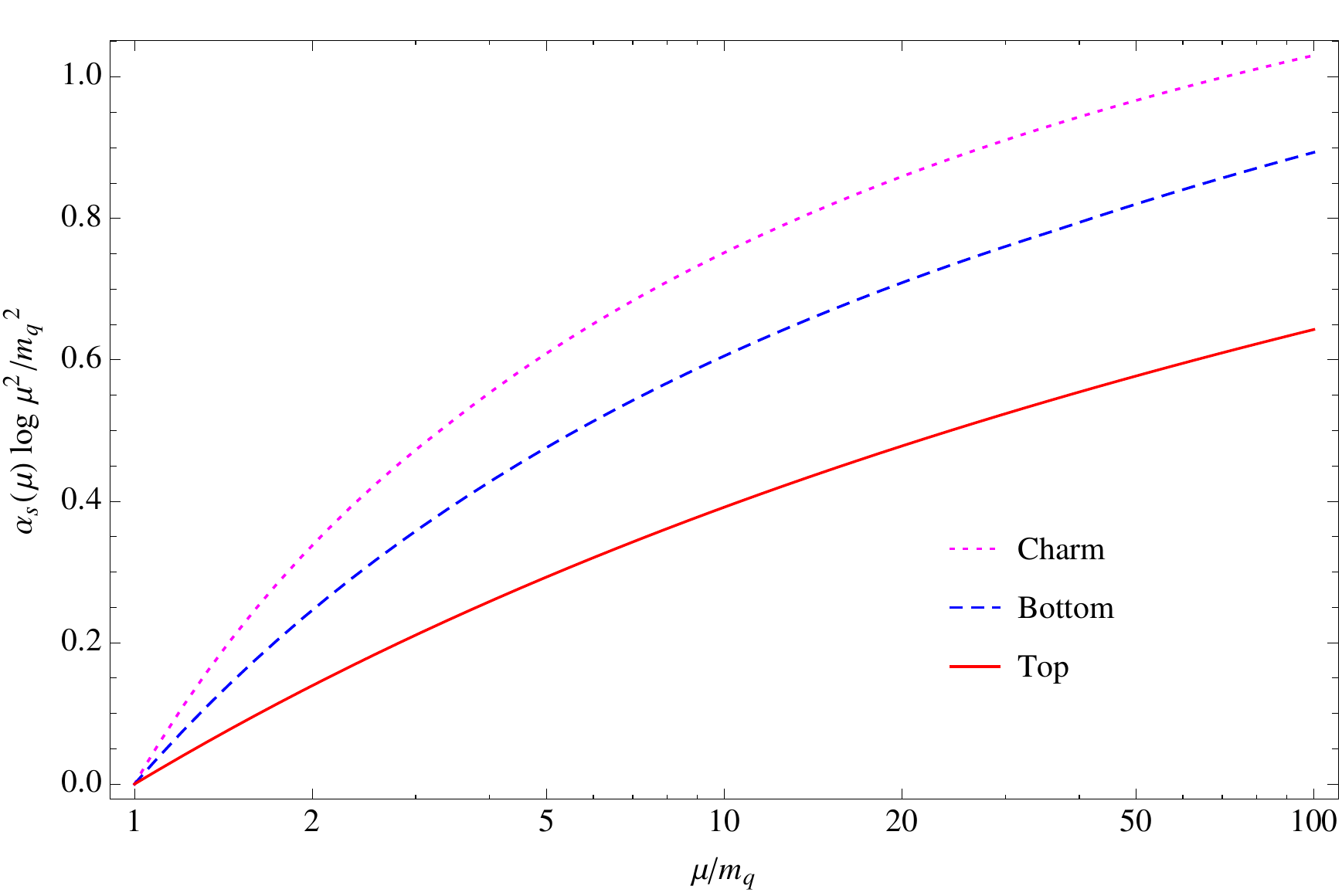}  
\caption{The size of $\alpha_s(\mu) \log \mu^2/m_q^2$ over the range $m_q\le
\mu \le 100 m_q$ for $q=c, b, t$. We use NLO running of $\alpha_s$ with
$\alpha_s(m_Z)$=0.119, $m_c=1.41$ GeV, $m_b=4.75$ GeV, and $m_t=175$ GeV. 
}
\label{fg:fig2}
\end{figure}
As is evident from the discussion in the previous section, the most important
 factor regarding the necessity of resumming the initial state collinear logarithms
  is the size of $\alpha_s(\mu) \log \mu^2/m_q^2$. In this regard it is informative to consider the size of this logarithm for the charm, bottom and top quarks, which we plot in Fig.~\ref{fg:fig2}. We see that $\alpha_s(\mu) \log \mu^2/m_q^2$ is significantly smaller at $\mu=100\times m_q$ for $q=t$ than for $q=c, b$:
\bea
\alpha_s(100\times m_c) \log (10^4) &\sim& 1.02 \  \nonumber \\  
\alpha_s(100\times m_b) \log (10^4) &\sim& 0.89 \  \\  
\alpha_s(100\times m_t) \log (10^4) &\sim& 0.64 \, , \  \nonumber
\eea
where we use $m_c=1.41$ GeV, $m_b=4.75$ GeV, and $m_t=175$ GeV.  Ref.~\cite{Maltoni:2012pa} studied the impact of including a  bottom quark PDF in order to resum collinear logarithms and found significant 
differences between the fully evolved $b$ quark PDF, $f_b(x,\mu)$, and the perturbative
approximations, ${\tilde f}_b^{(1),(2)}(x,\mu)$.  This difference is significantly smaller in the case of the
charm quark \cite{Buza:1996wv,Alekhin:2009ni}.
We see that the reason is simply the  asymptotic freedom of QCD, which implies an even smaller effect from
resumming logarithms into
 a  top quark PDF.

\begin{figure}[t]
\subfloat[LO Ratio]{\includegraphics[scale=0.6, angle=0]{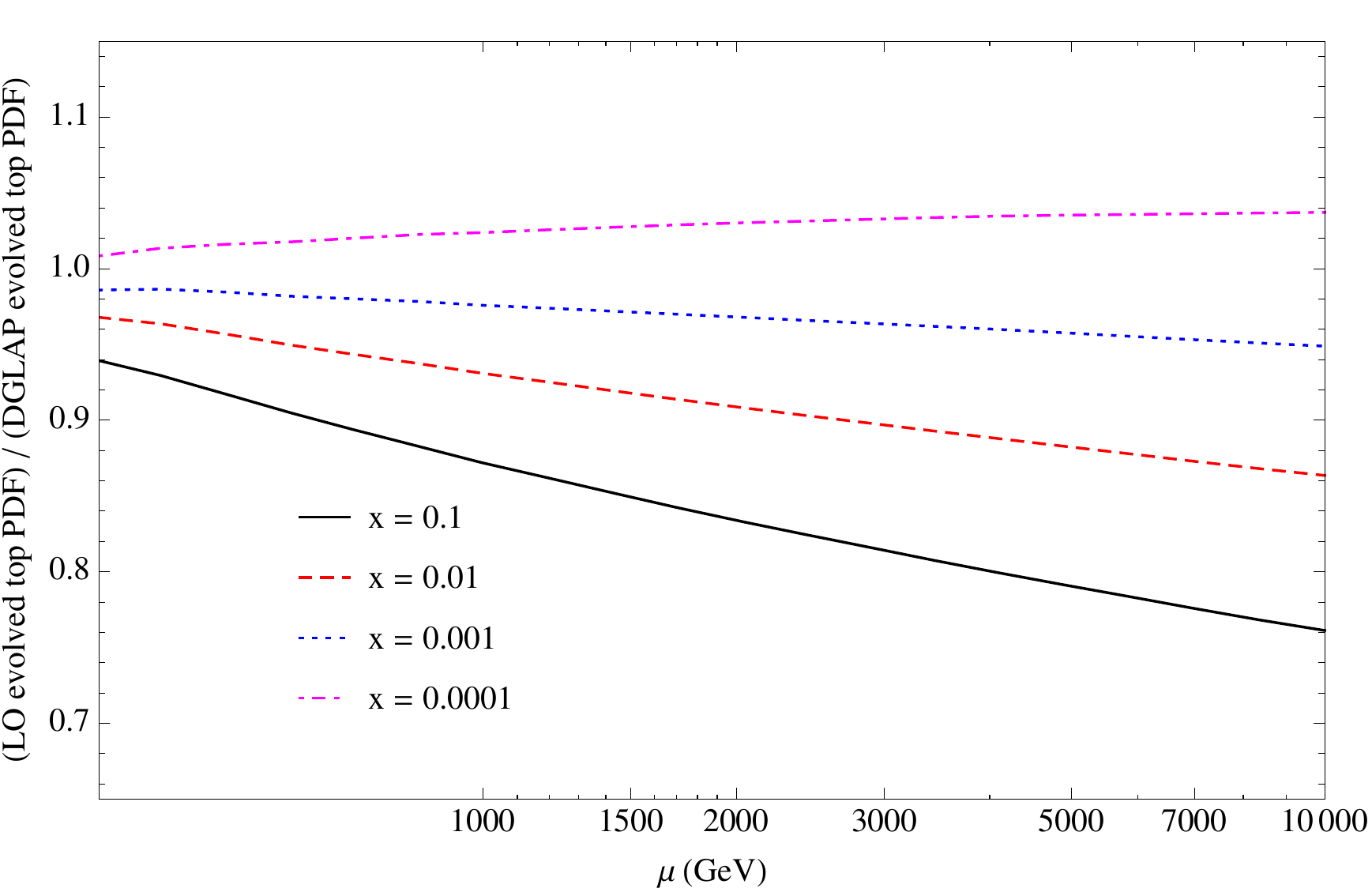}\label{fg:toplo}} \\
\subfloat[NLO Ratio]{\includegraphics[scale=0.6, angle=0]{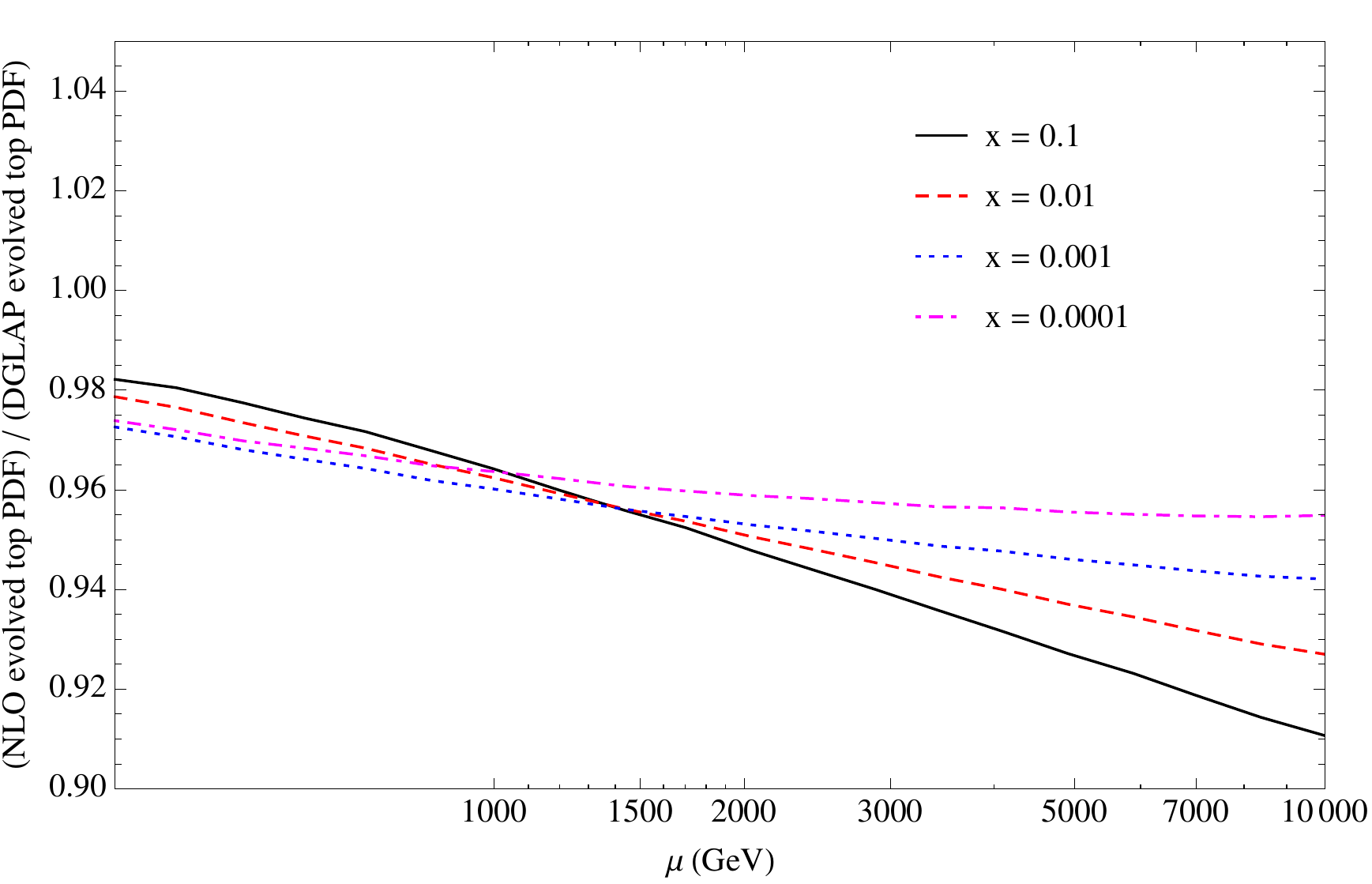}\label{fg:topnlo}}
\caption{Ratios of the perturbatively evolved top quark PDF, ${\tilde f}_t(x,\mu)$,  to the solution of the DGLAP 
equations for the top quark PDF, $f_t(x,\mu)$, at LO and NLO. In (a) we show the ratio using ${\tilde f}^{(1)}_t(x,\mu)$ and the LO NNPDF set {\tt NNPDF23\_lo\_as\_0119}~\cite{Ball:2012cx}. In (b) we show the ratio using ${\tilde f}^{(2)}_t(x,\mu)$ and the NLO NNPDF set {\tt NNPDF23\_nlo\_as\_0119}~\cite{Ball:2012cx}.}
\end{figure}

To evaluate the impact of resumming the logarithms in the 
case of the 
top  quark PDF explicitly, we follow Ref.~\cite{Maltoni:2012pa} and plot the ratio $\tilde{f}_t(x,\mu)/f_t(x,\mu)$, where $\tilde{f}_t(x,\mu)$ are the perturbative PDFs defined in Eqs.~(\ref{eq:1looppdf}) and (\ref{eq:2looppdf}) and $f_t(x,\mu)$ are the DGLAP-evolved top PDFs at the corresponding perturbative order. In this ratio, we expect that the uncertainties in the gluon and light flavor PDFs should largely cancel. The comparison is shown in Fig.~\ref{fg:toplo} and Fig.~\ref{fg:topnlo} for different values of Bjorken $x$. We see that, at NLO, the difference between the 2-loop approximated PDF, ${\tilde f}^{(2)}_t(x,\mu)$ and the fully evolved PDF, $f_t(x,\mu)$ is very small, of the order of 5\% level unless one chooses very large $\mu\sim 10$ TeV.  From this we conclude that the
sub-dominant logarithms in the DGLAP equations are numerically small. 

Fig.~\ref{fg:toplo} and Fig.~\ref{fg:topnlo} also show the second factor affecting the impact
of resumming collinear logarithms into
 a top PDF, the Bjorken $x$. From the  figures we see that
 the effect of resummation is larger, relatively speaking, at larger $x$. This feature can be understood from the evolution equation for $f_q(x,\mu)$:
\be
\frac{d}{d\log\mu} f_q(x,\mu) = \frac{\alpha_s(\mu)}{\pi} \int_{x}^1 \frac{dy}{y} \left[ P_{qg}\left(\frac{y}x,\mu\right) f_g(x,\mu) + P_{qq}\left(\frac{y}x,\mu\right) f_q(x,\mu) \right] \ ,
\label{evolvef}
\ee
which is the simple statement that there are two possibilities to produce a heavy quark $q_h$ in the $n_f$ FNS, the splitting of a gluon into a $q_h\bar{q}_h$ pair and the splitting of  $q_h$ into $gq_h$. The gluon splitting function $P_{qg}$ was given in Eq.~(\ref{eq:pqg}) while the quark splitting function is \cite{Altarelli:1979ub}
\be
\label{eq:pqq}
P_{qq}(z) = \frac{4}3 \left[ \frac{1+z^2}{(1-z)_+} + \frac32 \delta(1-z)\right] \ ,
\ee
where the plus-distribution is defined as
\be
\int dz f(z) g(z)_+ = \int dz \left[ f(z) - f(1) \right] g(z) \ .
\ee
The important observation here is that
$P_{qq}$ in Eq.~(\ref{eq:pqq}) has a peak at $z=1$. Therefore,
 the contribution of $P_{qq}$ to the evolution of $f_t(x,\mu)$ is more important near $x=1$, resulting in a larger effect 
 from the  resummation
 of the logarithms implicit in Eq. (\ref{evolvef}).
  In a hadron collider at center-of-mass energy $\sqrt{S}$, 
  a new particle produced at the rapidity $y$ and mass $m_\phi$ probes the momentum fractions
\be
x_1 = \frac{m_\phi}{\sqrt{S}} e^{y} \ , \qquad x_2 = \frac{m_\phi}{\sqrt{S}} e^{-y}\ .
\ee 
So for production of a particle with some fixed mass $m_\phi$ at a given rapidity, a collider with a larger $\sqrt{S}$ would require
typically smaller values of $x$, where the top quark PDF is well approximated by the NLO perturbative result,
as can be seen explicitly from  Fig.~\ref{fg:topnlo}. That is,
 the perturbative expansion is expected to be more accurate at higher $\sqrt{S}$. 
 This is the same observation as  in $b$-quark initiated processes, such as  single top and $hb\bar{b}$ production, where effects of resumming logarithms into a  $b$ PDF are more pronounced at the Tevatron than at the LHC~\cite{Maltoni:2012pa}. 

The third factor is related to the fact that the hard momentum transfer, $Q^2$, although estimated to be of the same order as $m_\phi^2$, is in reality slightly less than $m_\phi^2$ due to phase space suppression. This is emphasized and demonstrated very clearly in Ref.~\cite{Maltoni:2012pa} in the case of the bottom quark. For 
the top quark the argument is no different. In the $(n_f-1)$ FNS, where the production is given by the 2-to-2 process $gq_l\to \phi q_h$,  the hard momentum transfer $Q^2$ is a dynamical scale set on an event-by-event basis as \cite{Maltoni:2012pa}
\be
Q^2(z) = m_\phi^2 \frac{(1-z)^2}{z} \ , \quad z = \frac{m_\phi^2}{\hat{s}} \ .
\ee
In other words, the cross section for  $gq_l\to \phi q_h$ in the $(n_f-1)$ FNS in the collinear region reproduces $q_l \bar{q}_h \to \phi$ convoluted not with Eq.~(\ref{eq:1looppdf}) using $\mu^2=m_\phi^2$, but with the following expression~\cite{Olness:1987ep,Barnett:1987jw,Maltoni:2012pa}:
\be
\frac{\alpha_s}{2\pi} \int_x^1 \frac{dz}z\ P_{qg}(z)\ f_g(x/z,\mu) \log\left(\frac{m_\phi^2}{m_q^2}\frac{(1-z)^2}{z}\right)\ .
\ee 
The argument of the logarithm is smaller than the simple ratio $m_\phi^2 / m_q^2$. More specifically, comparing $\sigma^{(n_f-1)}$ in the collinear region with $\sigma^{(n_f)}$ we have \cite{Olness:1987ep,Barnett:1987jw,Maltoni:2012pa},
\bea
\sigma^{(n_f-1)} &\to&  \hat{\sigma}(q_l {\overline q}_h\to \phi) \Bigg\{ \int_\tau^1 \frac{dx}{x} f_{q_l}^{(n_f-1)}\left(\frac{\tau}{x},\mu\right) \nonumber \\
&&  \quad \times \int_x^1 \frac{dz}{z}  \frac{\alpha_s}{2\pi} f_g^{(n_f-1)}\left(\frac{x}{z},\mu\right) P_{qg}(z) \log\left(\frac{m_\phi^2}{m_{q_h}^2}\frac{(1-z)^2}{z}\right) \nonumber \\
&& \ + \int_\tau^1 \frac{dx}{x} f_{q_l}^{(n_f-1)}\left(x,\mu\right) \nonumber \\
&& \quad \times \int_x^1 \frac{dz}{z}  \frac{\alpha_s}{2\pi} f_g^{(n_f-1)}\left(\frac{\tau}{xz},\mu\right) P_{qg}(z) \log\left(\frac{m_\phi^2}{m_{q_h}^2}\frac{(1-z)^2}{z}\right) \Bigg\} \ , \\
\sigma^{(n_f)} &\to&\hat{\sigma}(q_l {\overline  q}_h\to \phi)  \Bigg\{ \int_\tau^1 \frac{dx}{x}  f_{q_l}^{(n_f)}\left(\frac{\tau}{x},\mu\right) \int_x^1 \frac{dz}zf_{q_h}^{(n_f)}\left(\frac{x}{z},\mu\right) \nonumber \\
&& \ + \int_\tau^1 \frac{dx}{x}  f_{q_l}^{(n_f)}\left(x,\mu\right) \int_x^1 \frac{dz}zf_{q_h}^{(n_f)}\left(\frac{\tau}{xz},\mu\right) \Bigg\} \ ,
\eea
where $\tau= (m_q+m_\phi)^2/S$ and $\hat{\sigma}(q_l {\overline q}_h\to\phi)$ is the partonic cross-section for the 2-to-1 process.

Taken together, these arguments demonstrate that the effect of resumming collinear logarithms into a top quark PDF at a high energy hadron collider would be significantly smaller than one might typically expect, and indeed less important than that of resumming analogous logarithms into a bottom quark PDF at the LHC.

\section{An Example: The Charged Higgs Production}
\label{sec:charged}
As an example of the effect of the resummation of large logarithms into the top PDF, we now consider inclusive charged Higgs production. Charged Higgs production in association with a top and bottom quark has been considered, both at LO
and at NLO, previously in the 
literature~\cite{Barnett:1987jw,Olness:1987ep,Zhu:2001nt,Plehn:2002vy,Boos:2003yi,Berger:2003sm,Dittmaier:2009np,Weydert:2009vr}.  Here, we re-examine the rate at $\sqrt{S}=100~\mathrm{TeV}$ in a 6FNS and 
numerically assess the impact of resumming
collinear logarithms into a top quark PDF by comparing to a 5FNS calculation.
We consider a charged Higgs that couples with the $H^+ \bar{t} b$ vertex,
\begin{equation}
\Gamma = \frac{i g}{2 \sqrt{2}} \left( g_L (1 - \gamma^5) + g_R (1 + \gamma^5) \right)
\, .
\end{equation}
In our results below, we take the MSSM couplings
\be
g_L= {m_t\over m_W\tan\beta}\qquad \qquad g_R={m_b\tan\beta\over M_W}\, ,
\ee
with $\tan \beta = 5$ for illustration. We reproduce the relevant contributions to the charged Higgs cross section here for convenience.

\subsection{LO}
At LO in the 6FNS, there is only the
tree level contribution from $t {\overline b}\rightarrow H^+$,
\be
\sigma_0={\pi g^2\over 24 {\hat s}}(g_L^2+g_R^2)
\int _0^1 dx\biggl[f_t(x){f_{\overline{b}}}\biggl({\tau\over x}\biggr)+
f_t\biggl({\tau\over x}\biggr){f_{\overline{b}}}(x)\biggr]\, ,
\label{eqn:sig0}
\ee
where $t$ and $\bar{b}$ are considered as massless partons. In the language of the 5FNS, this is simply the leading log approximation to the full cross section.
It contains all terms $c_{n1} (\alpha_s L)^n$, and so is correct up to terms of order $\alpha_s$.

\subsection{Comparing to the 5FNS}

While the above calculation provides a better approximation to the full cross section than the 5FNS LO calculation when the collinear logarithm arising from gluon splitting is large, it is insufficient to describe charged Higgs production at finite $p_T$. In particular, the 5FNS LO calculation includes the process $g \bar{b} \to \bar{t} H^+$, which provides the leading contribution to the Higgs $p_T$ distribution. To compare our 6FNS calculation to the 5FNS, we now add this process to the 6FNS LO calculation. The spin- and color-averaged amplitude for charged Higgs production $g(p) \bar{b}(p') \to \bar{t}(k)  H^+(k')$ is given by \cite{Barnett:1987jw},
\begin{eqnarray}
\label{eqn:amplitude}
{\hat{\sigma}}_{gb}
&=&{1\over 16\pi {\hat s}^2}
\int_{t_\mathrm{min}}^{t_\mathrm{max}}
|\mathcal{M}|^2 dt
\nonumber \\
|\mathcal{M}|^2 &=& \frac{2 \pi^2 \alpha \alpha_S (g_L^2 + g_R^2)}
{3 \sin^2 \theta_W} \bigg( \frac{\h{s} - 2 m_t^2}{m_t^2 - \h{t}} + \frac{2 m_t^2 (m_H^2 - \h{t})}{(m_t^2 - \h{t})^2} 
\nonumber \\
&&
+ \frac{m_t^2 - \h{t}}{\h{s}}
 - \frac{2(m_H^2 - \h{t})(\h{s} + m_t^2 - m_H^2)}{\h{s} (m_t^2 - \h{t})} \bigg)\, ,
\end{eqnarray}
where 
$\h{s} = (p + p')^2$, $\h{t} = (p - k)^2$, $\h{u} = (p - k')^2$. 
The contribution to the hadronic cross section is,
\begin{equation}
\sigma_1=\int dx_1 dx_2 {\hat{\sigma}}_{gb}
({\hat{s}})
\biggl[
f_g(x_1)
{f_{\bar{b}}}(x_2)+
(1\leftrightarrow 2)\biggr]\, .
\label{gb22}
\ee
In this expression, $b$ is taken as a massless parton, while the top quark mass is retained, in
agreement with the S-ACOT scheme \cite{Kramer:2000hn}.

\begin{figure}
\includegraphics[scale=0.67, angle=0]{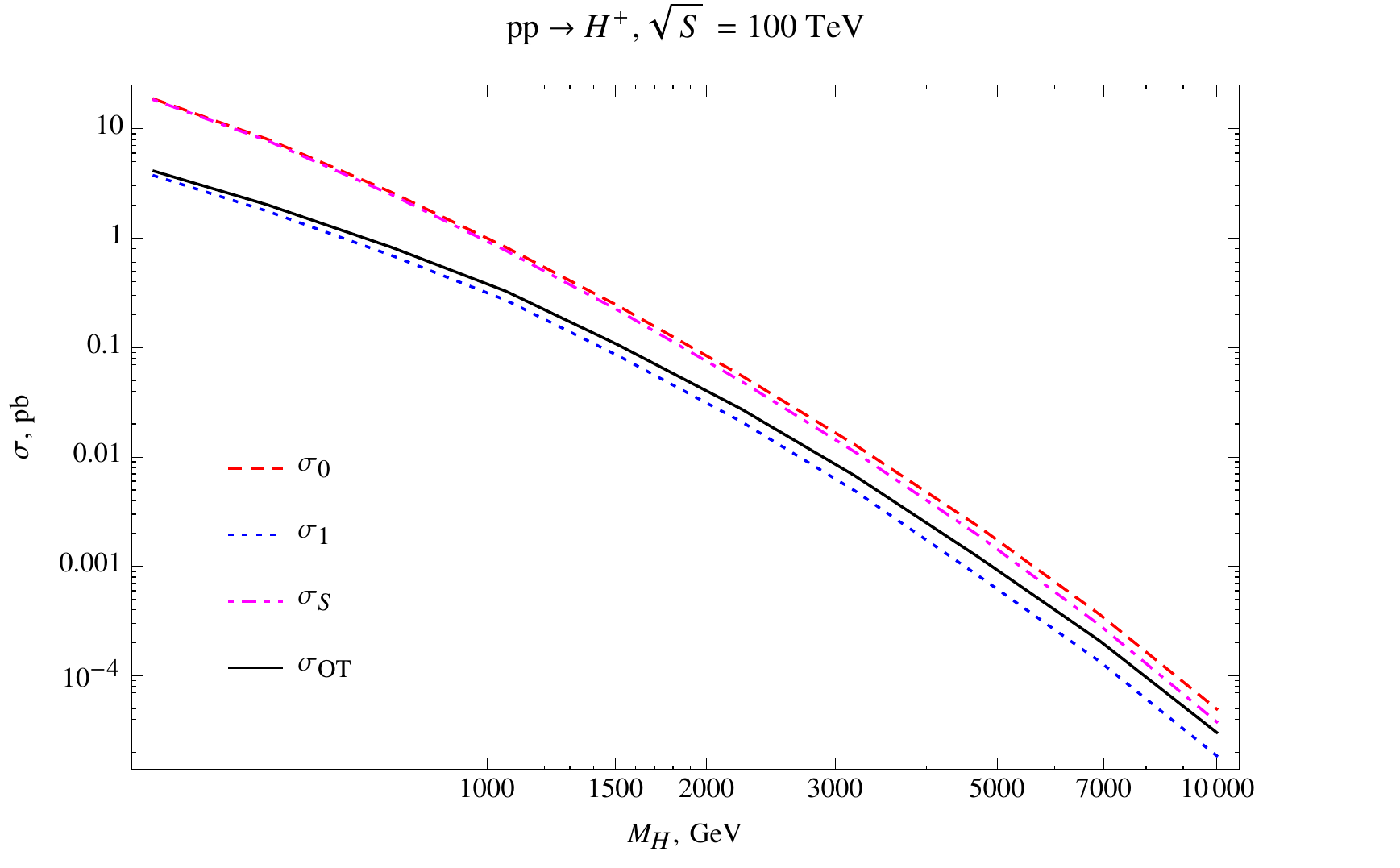}  
\caption[]{The calculation of the inclusive charged 
Higgs cross section in the 6FNS using Eq.~(\ref{sigot}). All curves use the PDF set {\tt NNPDF23\_lo\_as\_0119}.}
\label{fig:ot}
\end{figure}

Eq.~(\ref{gb22}) contains a contribution where the gluon splits into a collinear $t {\overline t}$ pair,
followed by the top quark scattering from the incoming $b$ quark,
\be
\label{eqn:subtraction}
\sigma_S=
{\pi g^2\over 24 {\hat s}}(g_L^2+g_R^2)
\int _0^1 dx\biggl[
{\tilde{f}_t^{(1)}}(x){f_{\bar{b}}}\biggl({\tau\over x}\biggr)+
{\tilde{f}_t^{(1)}}\biggl({\tau\over x}\biggr){f_{\bar{b}}}(x)\biggr]\, ,
\ee
where ${\tilde f}_t^{(1)}$ is the ${\cal O}(\alpha_s)$ perturbative approximation to the top quark distribution of Eq.~(\ref{eq:1looppdf}).
This contribution is already 
included in Eq.~(\ref{eqn:sig0}) and must be subtracted in order to avoid double counting. The consistent total cross section is
\be
\sigma_\mathrm{OT}(pp\rightarrow H^+ X)=\sigma_0+\sigma_1-\sigma_S\, ,
\label{sigot}
\ee
where the subscript indicates that this is the final result of the authors of~\cite{Olness:1987ep}. $\sigma_\mathrm{OT}$ contains all contributions of order $(\alpha_s L)^n$ and $\alpha_s$, and hence captures the LO + LL calculation of the 5FNS. In Fig.~\ref{fig:ot}, we show the three contributions along with the final result as a function of $m_H$,
for $\mu_F=\mu_R=m_H$, using the 6FNS LO PDF set \verb+NNPDF23_lo_as_0119+. The subtraction term nearly cancels against the LO cross section for small $m_H$. Even for large $m_H$ where one expects the logarithms to be large, the cancellation between the subtraction term and the LO cross section is still quite effective, signaling the effect of resumming logarithms into the top PDF to be small. At large $m_H$ the difference  between the $2\to2$ cross section and the full result is in the order of 50\%.

While $\sigma_0$ is significantly larger than $\sigma_1$, its influence is canceled nearly completely by $\sigma_S$. The relative difference between $\sigma_1$ and $\sigma_\mathrm{OT}$ corresponds to the effect of the $c_{21}, c_{31},\ldots$ terms in the cross section that are obtained in the 6FNS by using the top PDF. 
The difference
 is small up to very large charged Higgs masses, indicating that a fairly reliable prediction for charged Higgs production may be obtained from the 5FNS, where the leading process is $g {\overline b}\rightarrow  {\overline t} H^+$.

\subsection{NLL}

We now calculate the charged Higgs cross section at next-to-leading-logarithm order,
 including all terms in the first 2 columns of Eq.~(\ref{eq:signf1}) consistently. In order to capture the effect of these terms, we must refine the calculation of the previous section as follows:
\begin{itemize}
\item We employ 6NFS NLO PDFs.
\item We include real and virtual corrections to the 6NFS LO calculation.
\item We include the new process $g  t \to b  H^+$.
\end{itemize}
The first of these changes is straightforward, and our results below use the PDF set \verb+NNPDF23_nlo_as_0119+.\footnote{The S-ACOT scheme is equivalent to the FONLL-A scheme of the NNPDF
 collaboration~\cite{Binoth:2010ra} for the NLO PDF set~\cite{Ball:2011mu}.}

The real and virtual corrections to $t  \bar{b} \to H^+$ may be written \cite{Harlander:2003ai},
\bea
\h{\sigma}_0^{\alpha_s} &=& \h{\sigma}_0 \Bigg\{ \delta(1 - z) \left[ 1 - \frac{4 \alpha_s}{3 \pi} \left( 1 - \frac{\pi^2}{3} \right) \right] \nonumber \\
&&+ \frac{4 \alpha_s}{3 \pi} \left[ 1 - z + (1 + z^2) \left( \frac{\log (1 - z)^2}{(1 - z)} \right)_+ + \frac{1 + z^2}{(1-z)_+} \log \left(\frac{\h{s}}{\mu^2}\right) \right] \Bigg\} \quad,
\eea
where $\h{\sigma}_0 = {\pi g^2}(g_L^2+g_R^2)/(24 {\hat s})$ is the leading order partonic cross section and $z = {m_H^2}/{\h{s}}$. This cross section may be convoluted with the PDFs in the usual way to give the hadronic cross section $\sigma_0^{\alpha_s}$.

Finally, the cross section $\sigma'_1$ for $g  t \to b H^+$ is given by Eqs.~(\ref{eqn:amplitude})--(\ref{gb22}) with $t \leftrightarrow b$. Here, $t$ is taken as a massless parton, again in accordance with the S-ACOT scheme. The $b$ mass is retained, though its effect is minimal. Just as the expression for $\sigma_1$ contains a contribution from a gluon splitting into a collinear $t\bar{t}$ pair, $\sigma'_1$ contains a contribution from a gluon splitting into a collinear $b\bar{b}$ pair, and so we must subtract the double-counted term analogous to Eq.~(\ref{eqn:subtraction}) for consistency:
\be
\sigma'_S=
{\pi g^2\over 24 {\hat s}}(g_L^2+g_R^2)
\int _0^1 dx\biggl[
{f_t}(x){\tilde{f}_{\bar{b}}^{(1)}}\biggl({\tau\over x}\biggr)+
{f_t}\biggl({\tau\over x}\biggr){\tilde{f}_{\bar{b}}^{(1)}}(x)\biggr]\, ,
\ee
where now ${\tilde{f}_{\bar{b}}^{(1)}}(x)$ is the one-loop bottom PDF defined according to Eq.~(\ref{eq:1looppdf}).

\begin{figure}
\includegraphics[scale=0.67, angle=0]{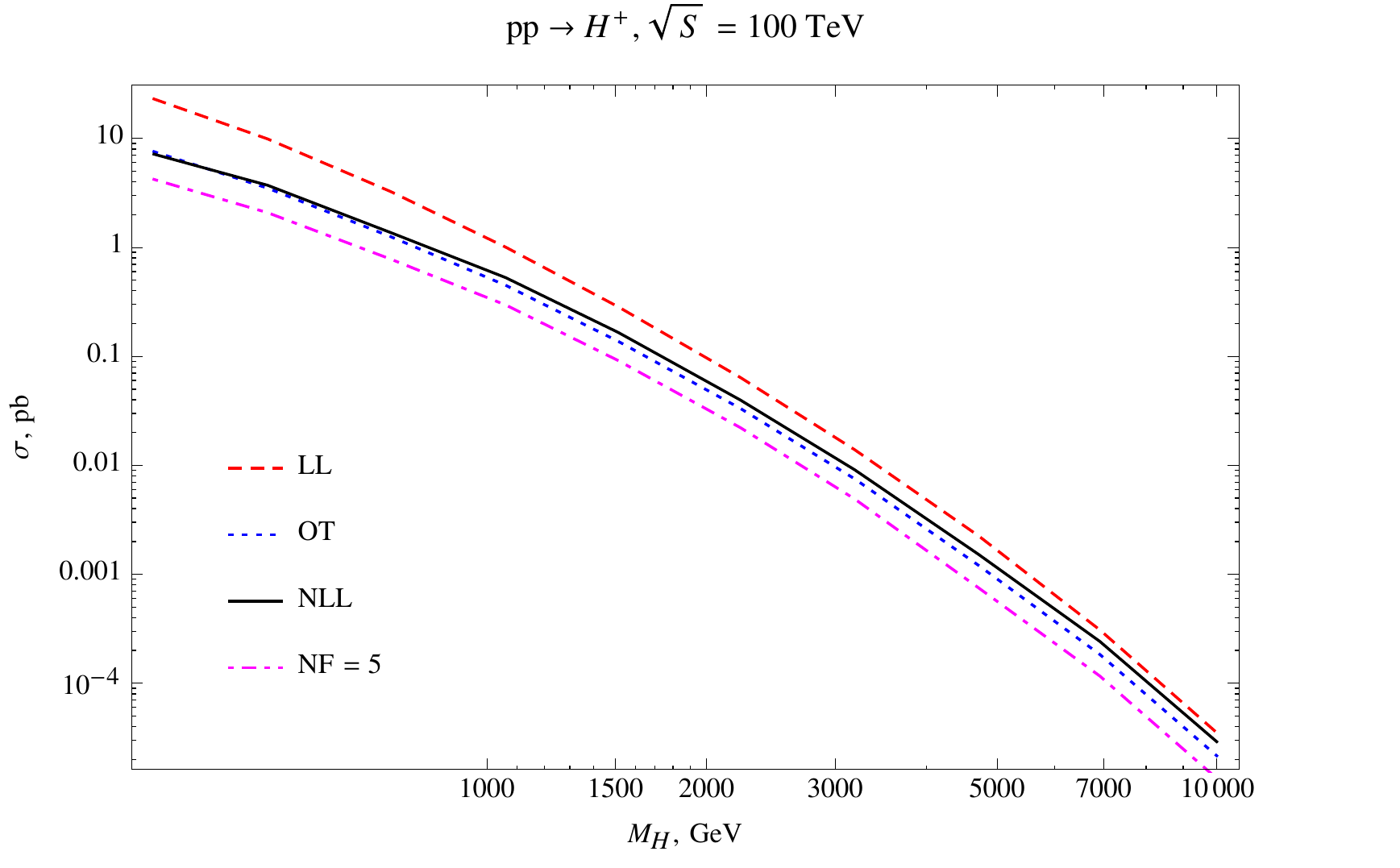}  
\caption[]{Comparison of the LL, LO + LL, and NLL cross sections for charged Higgs production. All 6FNS curves use the PDF set {\tt NNPDF23\_nlo\_as\_0119}, while the ``NF = 5'' curve uses the PDF set {\tt NNPDF23\_nlo\_FFN\_NF5\_as\_0119}.}
\label{fig:nll}
\end{figure}

Putting everything together, we have the full NLL cross section
\be
\sigma_\mathrm{NLL}(pp\rightarrow H^+ X)=\sigma_0^{\alpha_s}+\sigma_1-\sigma_S+\sigma'_1-\sigma'_S\, ,
\ee
which contains all terms proportional to $(\alpha_s L)^n$ and $\alpha_s (\alpha_s L)^n$.

\begin{figure}[t]
\subfloat[]{\includegraphics[scale=0.73, angle=0]{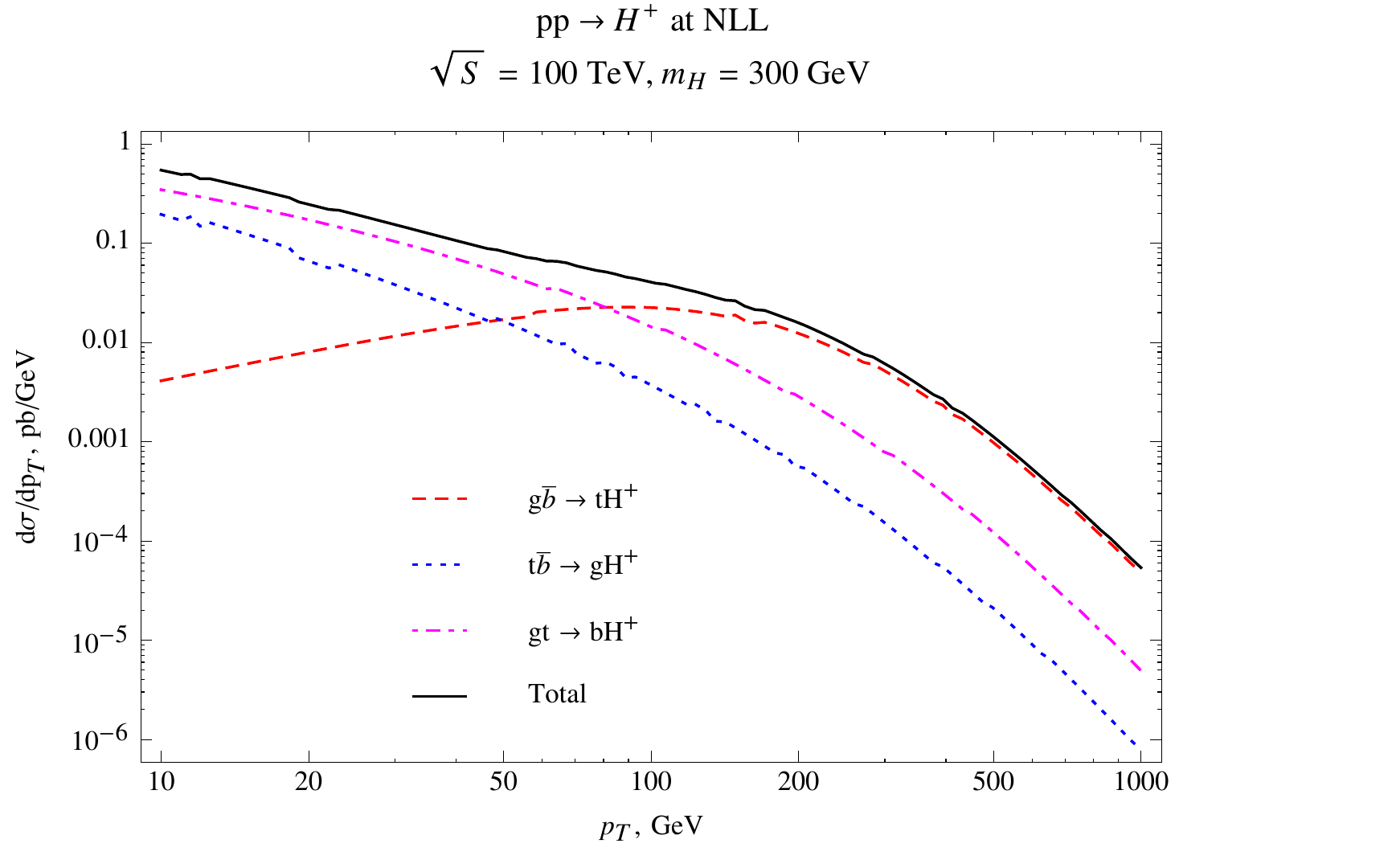}\label{fig:dsdpt}} \\
\subfloat[]{\includegraphics[scale=0.73, angle=0]{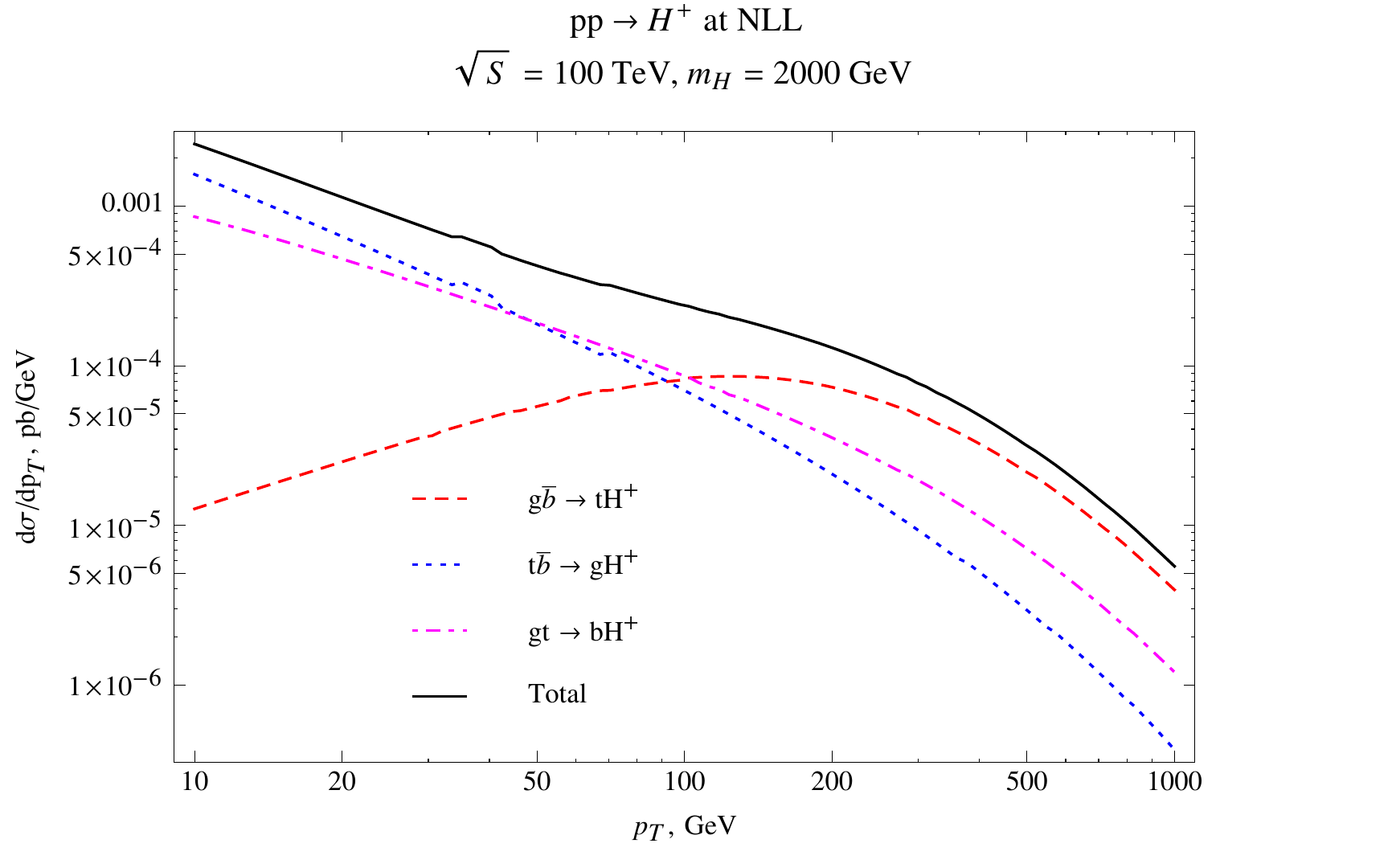}\label{fig:dsdpt2}}
\caption{$p_T$ distributions of the charged Higgs in various processes contributing to its production.}
\label{fig:fig6}
\end{figure}

The result of the full NLL calculation is compared with those of the previous sections in Fig.~\ref{fig:nll}. In this figure, we also show the
LO contribution in a 5FNS scheme from the partonic scattering $g{\overline b}
\rightarrow {\overline t} H^+$, using 5FNS PDFs.  While the relative impact of switching from LO to NLO PDFs is small, as we can see by comparing the LL curves in Figs.~\ref{fig:ot} and \ref{fig:nll}, the change to NLO PDFs significantly affects the cancelation between $\sigma_0$ and $\sigma_S$, as evidenced by the $\sigma_\mathrm{OT}$ curves in these figures. Moreover, the effects of including the new process $g  t \to b  H^+$ with the appropriate subtraction and  the QCD corrections to $t\bar{b} \to H^+$ nearly cancel, as the full NLL calculation is quite close to the LO + LL result. Our final NLL result varies from the LO 5NFS calculation by a factor of $\approx 2-3$, and this difference corresponds to the effect of including the cross section terms proportional to $(\alpha_s L)^{n + 1}$ and $\alpha_s (\alpha_s L)^n$ for $n \geq 1$ by resumming collinear logarithms into the top PDF.

We also study whether this cancellation occurs consistently over the range of kinematic variables.
In Figs.~\ref{fig:dsdpt} and \ref{fig:dsdpt2}, we plot  ${d\sigma/ dp_T} $ 
from the three $2 \to 2$ contributions to the full NLL cross section for $m_{H}=300~\mathrm{GeV}$ and $m_{H}=2~\mathrm{TeV}$, taking $\mu=m_H.$\footnote{Our general expression for the differential $p_T$ distribution may be found in the Appendix.} In the 6FNS, Higgses with small $p_T \alt m_t$ mostly come from  $gt\to bH^+$ for $m_H=300$ GeV and $tb \to g  H^+$ for $m_H=2$ TeV, while those with large $p_T \agt m_t$ are generated in the $gb \to t  H^+$ channel. This can be easily understood from the kinematics as the top is quite massive and the bottom and the gluon are effectively massless. Fig.~\ref{fig:fig6} also suggests that the LO result in 5FNS is not sufficient to describe the charged Higgs production in the small $p_T\alt m_t$. In this regime one should either switch to a 6FNS calculation at NLL order, as is done in this work, or proceed to the NLO calculation in the 5FNS, which is more involved than the NLL computation presented here. Alternatively, one could interpolate between the NLL 6FNS result at small $p_T$ and the LO 5FNS result at large $p_T$, switching over at $p_T\sim m_t$.

It is also interesting to contrast the situation with the associated production with a $b$ quark. In this case, our findings from Fig.~\ref{fig:fig6} indicate that, at LO in 4FNS, the $p_T$ spectrum produced by the 2-to-2 process should agree with the spectrum from the NLL calculation in the 5FNS across a wide range of $p_T$: $p_T\agt m_b$. In other words, the $p_T$ distributions in both schemes arise from the same 2-to-2 process in associated production with a $b$ quark, while in the case of top quark the $p_T$ distribution at $p_T\alt m_t$ is generated from processes in 6FNS that are not existent in the 5FNS, {\em i.e.} $gt\to bH^+$ and $tb\to gH^+$.

\begin{figure}[t]
\includegraphics[scale=0.59, angle=0]{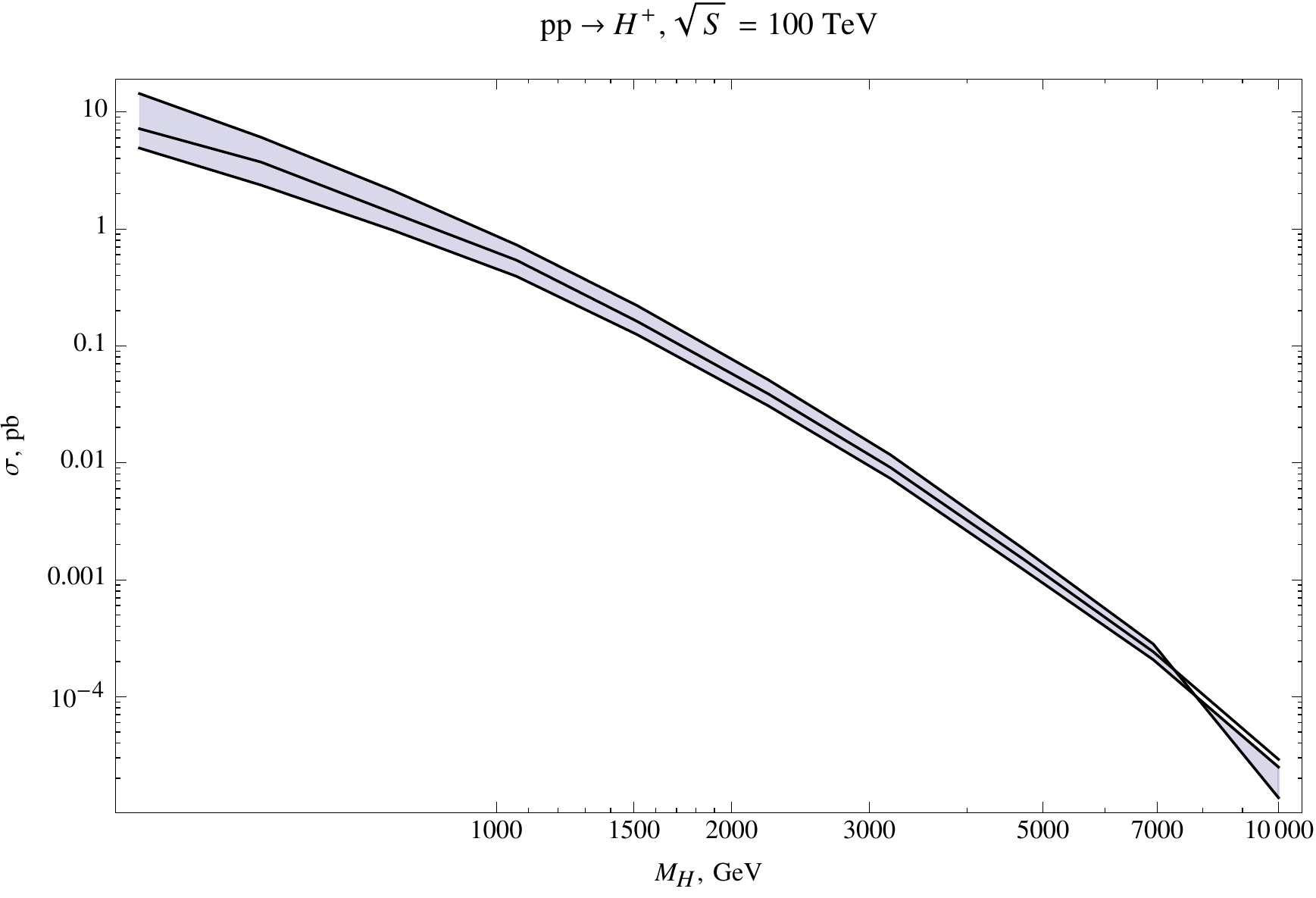}  
\caption[]{The scale dependence of the full NLL result for charged Higgs production. The three curves represent the cross section at scales $0.5, 1, 2 \times m_{H}$.}
\label{fig:scale}
\end{figure}

Finally the scale dependence is shown in Fig.~\ref{fig:scale}, where we show a band obtained by varying $m_H/2< \mu< 2 m_H$ in the strong coupling constant and all the PDFs entering the NLL cross section. The uncertainty corresponding to scale variation is considerable compared to the difference between the NLL result and the LO 5FNS cross section, suggesting good agreement between the 5FNS and the 6FNS cross sections.

\section{Conclusion}
\label{sec:conclusion}

In this work we studied the production cross section of a new heavy particle $\phi$ in association with a top quark in a high energy $pp$ collider. The collinear singularity in the cross section could be resummed into the top quark PDF, by treating the top as a parton inside the proton. This topic was first considered in Refs.~\cite{Barnett:1987jw,Olness:1987ep} before the discovery of the top quark. Given the relatively large mass for the top, we examined the necessity of introducing a top PDF in a future $pp$ collider at $\sqrt{S}=100$ TeV. Our findings suggest the effect of resummation of the collinear logs is, in general,  smaller than that in the case of associated production with the bottom or charm quark, for $m_\phi \alt 10$ TeV. In particular, including the perturbative expansion of the collinear logs to NLL in $\alpha_s$ turned out to be a very good approximation for the fully evolved NLO top PDF.

Using the production of a charged Higgs boson in the MSSM as an example, we computed the cross section at NLL in the 6FNS and compared with the LO cross section in the 5FNS. For the total cross section, we found good agreement between the LO 5FNS and the NLL 6FNS results, after taking into account the uncertainty resulting from the scale dependence. For the $p_T$ distribution, however, our computation indicates that the 5FNS distribution matches well with the 6FNS result only in the region of $p_T\gg m_t$. At $p_T\ll m_t$ the LO 5FNS result was significantly smaller than the NLL 6FNS because of the large $m_t$ in the final state, which suggests a NLO 5FNS calculation is needed in this regime. Alternatively, one could also interpolate between the 6FNS computation at $p_T\alt m_t$ and the 5FNS computation at $p_T\agt m_t$. This is in contrast with the associated production with a $b$ quark. Since $m_b$ is so small, the LO 4FNS result should already be able to generate a $p_T$ distribution for a wide range of $p_T$.

One topic we have not studied in this work is the inclusion of finite $m_t$ effects in the top PDF. They are important only in the region $Q^2 \sim m_t^2$, where the collinear logarithms are expected to be small. However, once a discovery is made in the future, precision measurements would require quantitative understanding of the finite $m_t$ effects. We hope to return to this issue in a future work.

\begin{acknowledgments}
S.D. thanks M.~Ubiali for a useful discussion of the FONLL scheme used by the
NNPDF collaboration.
The work of S.D. is supported by the U.S.~Department of Energy under grant
No.~DE-AC02-98CH10886. Work at ANL is supported by the U.S.~Department of
Energy under grant No.~DE-AC02-06CH11357. A.I. is supported in part by the U.S.~Department of
Energy under grant  DE-FG02-12ER41811. I.L. is supported in part by the U.S.~Department of Energy under grant No. DE-SC0010143.
\end{acknowledgments}

\appendix*
\section{Kinematics}
\label{kinematics}
Here, we review the calculation of the $p_T$ distribution for an arbitrary $2 \to 2$ process with massive particles in the final state. The initial particles are assumed to be massless.

For $p_a, p_b \to p_1, p_2$ the partonic cross section from Peskin and Schroeder is~\cite{Peskin:1995ev}
\begin{eqnarray}
\h{\sigma} &=& \frac{1}{(x_1 \sqrt{S}) (x_2 \sqrt{S}) |v_a - v_b|} \int \frac{1}{(2\pi)^6} \frac{d^3 p_1}{2 E_1} \frac{d^3 p_2}{2 E_2} (2\pi)^4 \delta^4(p_a + p_b - p_1 - p_2) |\mathcal{M}|^2 \nonumber \\
&=& \frac{1}{2\h{s}} \int \frac{1}{(2\pi)^3} \frac{d^3 p_1}{2 E_1} \frac{1}{2 E_2} (2\pi) \delta(\sqrt{\h{s}} - E_1 - E_2) |\mathcal{M}|^2 \nonumber \\
&=& \frac{1}{2\h{s}} \int \frac{1}{(2\pi)^3} \frac{dp_z p_T dp_T d\phi}{2 E_1} \frac{1}{2 E_2} (2\pi) \delta(\sqrt{\h{s}} - E_1 - E_2) |\mathcal{M}|^2 \nonumber \\
&=& \frac{1}{32\pi \h{s}} \int \frac{dp_z dp_T^2}{E_1 E_2} \delta(\sqrt{\h{s}} - E_1 - E_2) |\mathcal{M}|^2
\end{eqnarray}
where $S$ is the hadronic CM energy squared, $x_1$ and $x_2$ are the momentum fractions of partons $p_a$ and $p_b$ such that the partonic CM energy squared is $\h{s} = x_1 x_2 S$, $|\mathcal{M}|^2$ is the spin- and color-averaged amplitude, $p_z$ is the longitudinal momentum of particle 1, which may be positive or negative, and $p_T$ is the magnitude of the transverse momentum of either particle, which is always positive. All kinematic quantities are assumed to be in the partonic center of mass frame. The delta function may be written
\begin{eqnarray}
\delta(\sqrt{\h{s}} - E_1 - E_2) &=& \delta \left( \sqrt{\h{s}} - \sqrt{(p_z)^2 + (p_T)^2 + m_1^2} - \sqrt{(p_z)^2 + (p_T)^2 + m_2^2} \right) \nonumber \\
&=& \delta(f(p_z))
\end{eqnarray}
where
\begin{eqnarray}
f(p_z) &=& \sqrt{\h{s}} - \sqrt{(p_z)^2 + (p_T)^2 + m_1^2} - \sqrt{(p_z)^2 + (p_T)^2 + m_2^2} \\
\frac{df}{dp_z} &=& -\frac{p_z}{E_1} - \frac{p_z}{E_2}
\end{eqnarray}
Performing the $p_z$ integral yields
\begin{eqnarray}
\h{\sigma} &=& \frac{1}{32\pi \h{s}} \int \frac{dp_T^2}{E_1 E_2} \left( \frac{|p_z|}{E_1} + \frac{|p_z|}{E_2} \right)^{-1} |\mathcal{M}|^2 \\
\frac{d\h{\sigma}}{dp_T^2} &=& \frac{1}{32\pi \h{s}} \frac{1}{E_1 E_2} \left( \frac{|p_z|}{E_1} + \frac{|p_z|}{E_2} \right)^{-1} |\mathcal{M}|^2 \nonumber \\
&=& \frac{1}{32\pi \h{s}^{3/2} |p_z|} |\mathcal{M}|^2
\end{eqnarray}
where we must sum over \emph{both} the positive and negative solutions of $f(p_z) = 0$, that is $p_z = |p_z|$ and $p_z = -|p_z|$. In the CM frame,
\begin{eqnarray}
p_a &=& (\sqrt{\h{s}}/2, 0, 0, \sqrt{\h{s}}/2) \\
p_b &=& (\sqrt{\h{s}}/2, 0, 0, -\sqrt{\h{s}}/2) \\
p_1 &=& (E_1, p_T \cos \phi, p_T \sin \phi, p_z) \\
p_2 &=& (E_2, -p_T \cos \phi, -p_T \sin \phi, -p_z)
\end{eqnarray}
and for the positive solution,
\begin{eqnarray}
\h{t} &=& (p_a - p_1)^2 \nonumber \\
&=& (\sqrt{\h{s}}/2)(E_1 - |p_z|) \equiv \hat{t}_-\\
\h{u} &=& (p_b - p_1)^2 \nonumber \\
&=& (\sqrt{\h{s}}/2)(E_1 + |p_z|) \equiv \hat{t}_+
\end{eqnarray}
Keeping terms from both $p_z$ solutions, then, it is often convenient to write
\begin{eqnarray}
\frac{d\h{\sigma}}{dp_T^2} &=& \frac{1}{32\pi \h{s}^{3/2} |p_z|} \Big[ |\mathcal{M}|^2 \left( \h{s}, \h{t} = \hat{t}_-, \h{u} = \hat{t}_+\right) +|\mathcal{M}|^2 \left( \h{s}, \h{t} =\h{t}_+ , \h{u} = \h{t}_- \right) \Big] \ .
\end{eqnarray}
We may express the kinematic variables in the above cross section as
\begin{eqnarray}
E_1 &=& \frac{\h{s} + m_1^2 - m_2^2}{2 \sqrt{\h{s}}} \\
|p_z| &=& \sqrt{\frac{\lambda(\h{s}, m_1^2, m_2^2)}{4\h{s}} - p_T^2}\ ,
\end{eqnarray}
where $\lambda(a, b, c) = a^2 + b^2 + c^2 - 2ab - 2ac - 2bc$.
Then, for two colliding hadrons $A$ and $B$, the hadronic differential cross section for $P_A, P_B \to p_1, p_2$ is
\begin{equation}
\frac{d\sigma}{dp_T^2} = \sum_{a,b} \int dx_1 dx_2 f_{a/A}(x_1) f_{b/B}(x_2) \frac{d\h{\sigma}(p_a, p_b \to p_1, p_2)}{dp_T^2}
\end{equation}
where the sum runs over all partons $a,b$ with $p_a, p_b$ defined as above, and the integration runs over the region
\begin{equation}
\frac{(m_1^2 + p_T^2) + (m_2^2 + p_T^2)}{S} < x_1 x_2 < 1\ .
\end{equation}

\newpage
\bibliographystyle{JHEP}
\bibliography{top}

\end{document}